\documentclass[useAMS, usenatbib, onecolumn]{mn2e}
\usepackage{graphicx}
\usepackage{amssymb}

\newcommand\be{\begin{equation}}
\newcommand\ee{\end{equation}}
\newcommand\ba{\begin{eqnarray}}
\newcommand\ea{\end{eqnarray}}

\newcommand\bb[1] {\mbox{\boldmath{$#1$}}}
\newcommand{\Alfven}{Alfv\'{e}n~}
\newcommand\tomega{{\tilde\omega}}
\def\go{\mathrel{\raise.3ex\hbox{$>$}\mkern-14mu
             \lower0.6ex\hbox{$\sim$}}}
\def\lo{\mathrel{\raise.3ex\hbox{$<$}\mkern-14mu
             \lower0.6ex\hbox{$\sim$}}}

\title[Papaloizou-Pringle Instability of Magnetized Accretion Tori]
{Papaloizou-Pringle Instability of Magnetized Accretion Tori}

\author[W.~Fu \& D.~Lai]
{Wen Fu$^1$\thanks{Email: wenfu@astro.cornell.edu;
dong@astro.cornell.edu} and Dong Lai$^{1,2}$\footnotemark[1]\\
$^1$Department of Astronomy, Cornell University, Ithaca, NY 14853, USA\\
$^2$KITP, University of California, Santa Barbara, CA 93106, USA\\}

\pagerange{\pageref{firstpage}--\pageref{lastpage}} \pubyear{2010}

\begin{document}

\label{firstpage}
\maketitle

\begin{abstract}
Hot accretion tori around a compact object are known to be susceptible
to a global hydrodynamical instability, the so-called
Papaloizou-Pringle (PP) instability, arising from the interaction of
non-axisymmetric waves across the corotation radius, where the wave
pattern speed matches the fluid rotation rate.  However, accretion
tori produced in various astrophysical situations (e.g., collapsars
and neutron star binary mergers) are likely to be highly magnetized. We
study the effect of magnetic fields on the PP instability in
incompressible tori with various magnetic strengths and structures. In
general, toroidal magnetic fields have significant effects on the PP
instability: For thin tori (with the fractional width relative to the
outer torus radius much less than unity), the instability is suppressed at
large field strengths with the corresponding toroidal \Alfven speed $v_{A\phi}\go 0.2r\Omega$ (where $\Omega$ is the flow rotation rate). For thicker tori (with the fractional width of order 0.4 or larger), which are hydrodynamically stable, the
instability sets in for sufficiently strong magnetic fields (with $v_{A\phi}\go 0.2 r\Omega$).
Our results suggest that highly magnetized accretion tori may be
subjected to global instability even when it is stable against the usual
magneto-rotational instability.
\end{abstract}


%%%%%%%%%%%%%%%%%%%%%%%%%%%%%%%%%%%%%%%%%%
\section{Introduction}

Differentially rotating fluid flows, ubiquitous in astrophysics, can
exhibit rich dynamical behaviors. Papaloizou \& Pringle (1984)
discovered that accretion tori can be subjected to a global
non-axisymmetric instability that grows on a dynamical time-scale.
Accretion tori are bagel-shaped discs with high internal temperatures
and well-defined boundaries. They may be representative of certain
stages or regions of the inner accretion flows around black holes,
such as those found in active galactic nuclei and quasars (e.g.,
Begelman, Blandford \& Rees 1984). They may also form in the
gravitational collapse of the rotating core of massive stars (e.g.,
Woosley 1993) and after the merger of compact neutron star and black
hole binaries (e.g., Duez et al.~2009; Etienne et al.~2009; Rezzolla et al.~2010; Montero
et al.~2010), and thus are thought to be the central engine of
gamma-ray bursts (e.g., Meszaros 2006).
The Papaloizou-Pringle (PP) instability arises from the interaction
between non-axisymmetric waves across the corotation radius ($r_{\rm c}$), where the
wave pattern rotation frequency $\Omega_{\rm p}$ equals the background fluid
rotation rate $\Omega$ (e.g., Blaes \& Glatzel 1986; Goldreich, Goodman \&
Narayan 1986; Glatzel 1987b).
Waves outside the corotation radius ($r>r_{\rm c}$) have
$\Omega_{\rm p}$ larger than $\Omega(r)$ and carry positive energy, while
waves at $r<r_{\rm c}$ have $\Omega_{\rm p}<\Omega(r)$ and carry negative energy.
Instability occurs when the negative-energy waves inside $r_{\rm c}$
lose energy to the positive-energy waves outside $r_{\rm c}$, leading the
amplification of the wave amplitudes. To maintain the instability,
the waves must be efficiently reflected at the inner
and outer boundaries so that they are trapped in the torus.
The growth rate of the PP instability is maximal for a constant-angular
momentum torus. For a very thin torus (with the inner and outer radii close
to each other), the instability disappears when
$p=d\ln\Omega/d\ln r>-\sqrt{3}$; for a wider torus, the instability
persists (with decreasing growth rate) as the Keplerian
rotation profile ($p=-3/2$) is approached (e.g., Papaloizou \& Pringle 1985, 1987;
Goldreich et al.~1986; Zurek \& Benz 1986; Sekiya \& Miyama 1988).
Other properties of the PP instability, such as its
connection with the instability of vortices
(e.g., Glatzel 1987a), its non-linear evolution
(e.g., Goodman, Narayan \& Goldreich 1987; Hawley 1991) and the effect
of accretion (Blaes 1987), have been studied.

Interest in the PP instability waned in the 1990s when
Balbus \& Hawley (1991) pointed out that the Magneto-Rotational Instability
(MRI), originally studied for magnetized Taylor-Couette flows
(Velikhov 1959; Chandrasekhar 1961), can be important for astrophysical
accretion discs. Since the MRI is robust and requires only a weak magnetic field,
the nonlinear development of MRI may lead to efficient angular momentum
transport in accretion discs. Over the last two decades, numerous studies
have been devoted to the MRI and related issues such as MHD turbulence in the
disc (see, e.g., Balbus \& Hawley 1998 and Balbus 2003 for reviews;
a sample of recent numerical studies include
Hirose et al.~2009; Guan et al.~2009; Simon et al.~2009;
Davis et al.~2010; Fromang 2010; Longaretti \& Lesur 2010;
Sorathia et al.~2010).

Nevertheless, the question remains as to what happens to the original
PP instability in an accretion torus when a finite magnetic field is
present.
After all, the tori produced in various astrophysical situations
(e.g., binary mergers; Rezzolla et al.~2010; Montero et al.~2010) are
expected to be highly magnetized.
One might dismiss this question as purely academic since such
a magnetic torus is likely MRI unstable and therefore turbulent.  We
note, however, that the usual MRI operates on perturbations with
vertical structure (i.e., with finite vertical wavenumber $k_z$),
while the PP instability operates on perturbations with $k_z=0$. That
is, the PP instability pertains to the height-averaged behavior of
the disc. Therefore one might expect that the PP instability will
continue to operate even in the presence of MRI-induced
turbulence. Furthermore, in connection with Galactic black-hole X-ray
binaries, it has been suggested that accretion tori can support
discrete, trapped oscillation modes, which might explain
high-frequency quasi-periodic oscillations (e.g.,
Strohmayer 2001; Remillard \& McClintock 2006)
observed in a number of X-ray binary systems
(e.g., Rezzolla et al.~2003; Lee et al.~2004;
Schnittman \& Rezzolla 2006; Blaes et al.~2007; Sramkova et al.~2007; Montero et al.~2007).
Although it is currently not clear that pressure-supported tori provide a
realistic model for the accretion flow around a black hole in any
spectral state, structures resembling pressure-supported
tori do appear to be present in some non-radiative global MHD
simulations of MRI-driven turbulent accretion flows (e.g., Hawley \&
Balbus 2002; De Villiers et al.~2003; Machida et al.~2006).

There have been a number of previous studies on global MHD
instabilities in accretion flows. For example, Knobloch (1992), Kumar
et al.~(1994), Curry, Pudritz \& Sutherland (1994) and Curry \&
Pudritz (1995) carried out global analysis for the axisymmetric modes
with finite $k_z$ in differentially rotating flows threaded by
vertical and/or azimuthal magnetic fields, thus establishing the
robustness of MRI in these flows. Ogilvie \& Pringle (1996) studied
the non-axisymmetric instability of a cylindrical flow containing an
azimuthal field, while Curry \& Pudritz (1996) studied a similar flow
containing a vertical field. Both studies focused on modes with finite
vertical wavenumbers, which inevitably invite MRI. Although the effect
of boundaries is emphasized, a somewhat arbitrary rigid boundary
condition was adopted in these studies.
As far we are aware, the behavior of the PP instability for finite tori
with magnetic (both vertical and azimuthal) fields has not been
clarified.

In this paper, as part of our ongoing investigation of global oscillation
modes and instabilities of rotating astrophysical flows
(Tsang \& Lai 2008, 2009a, b; Lai \& Tsang 2009; Fu \& Lai 2010), we
carry out global stability analysis of magnetized accretion
tori subjected to nonaxisymmetric perturbations. Since our main aim is to
understand the effects of magnetic fields on the original PP
instability, we focus on modes with no vertical structure ($k_z=0$)
and we pay particular attention to the boundary conditions.
As in many previous studies mentioned above, we model the torus by
a cylindrical incompressible flow threaded
by both vertical and toroidal magnetic fields.

Our paper is organized as follows. In section 2, we describe the
equilibrium models for our rotating magnetized flows.  In section 3,
the basic perturbation equations are presented. We derive the boundary
conditions in section 4 and present our numerical calculations of the
global instability in section 5. Final summary and discussion of our
results are given in section 6.

%%%%%%%%%%%%%%%%%%%%%%%%%%%%%%%%%%%%%%%%%%%%%%%%%%%%%%%%%%%%%%%%%%
\section{Equilibrium Models}

As mentioned above, the PP instability operates in modes with no
vertical structure ($k_z=0$). As such, the dynamics can be captured by
height-averaged fluid equations. We consider a cylindrical shell (of
finite width) of incompressible non-self-gravitating fluid, which is
rotating differentially in the external gravitational field produced by a
central compact object. We adopt the cylindrical coordinates ($r,~\phi,~z$)
with the $z$-axis being the rotation axis. The cylindrical shell is
assumed to be infinitely long in the $z$-direction and
threaded by magnetic fields.
The fluid satisfies the ideal MHD equations:
\be
{\partial{\bb{u}} \over \partial t}
+({\bb{u}}\cdot\nabla){\bb{u}} =-\frac{1}{\rho}\nabla \Pi-{\nabla
  \Phi}+\frac{1}{4\pi\rho}(\bb{B} \cdot \nabla)\bb{B},\label{eq:1} \ee
\be
{{\partial {\bb{B}}} \over \partial t}=\nabla\times({\bb{u}}
\times{\bb{B}}),\label{eq:2} \ee
\be
\nabla \cdot\bb{B}=0, \label{eq:3}
\ee
\be \nabla \cdot \bb{u}=0.\label{eq:4} \ee
Here, $\rho$ is the constant fluid density, $\bb{u}$ the fluid
velocity, $\bb{B}$ the magnetic field,
and $\Pi=P+\bb{B}^2/8\pi$ the total pressure with $P$ being the
gas pressure. The gravitational potential is $\Phi=-GM/r$, where
$M$ is the mass of the central object. The background flow velocity
$\bb{u}=r\Omega(r)\hat{\bb{\phi}}$
and magnetic field $\bb{B}=B_{\phi}(r)\hat{\bb{\phi}}+B_z(r)\hat{\bb{z}}$ also depends
only on $r$. For convenience, we assume the flow has a power-law rotation
profile
\be
\Omega(r) \propto r^p.
\ee

The flow is confined between two boundaries ($r_1 \le r \le r_2$), where the
gas pressure vanishes ($P|_{r_1,~r_2}=0$). Outside the fluid zone is a
vacuum devoid of matter but maybe permeated with magnetic fields.
In the equilibrium state, we assume that the magnetic field is continuous across
the fluid boundaries so that there is no surface electric current at $r=r_1,\,r_2$
(however, we allow for surface current to develop when the fluid is perturbed).
We will consider two models of magnetic field structure.

\subsection{Model (a)}
\begin{figure}
\begin{center}
\includegraphics[width=0.6\textwidth]{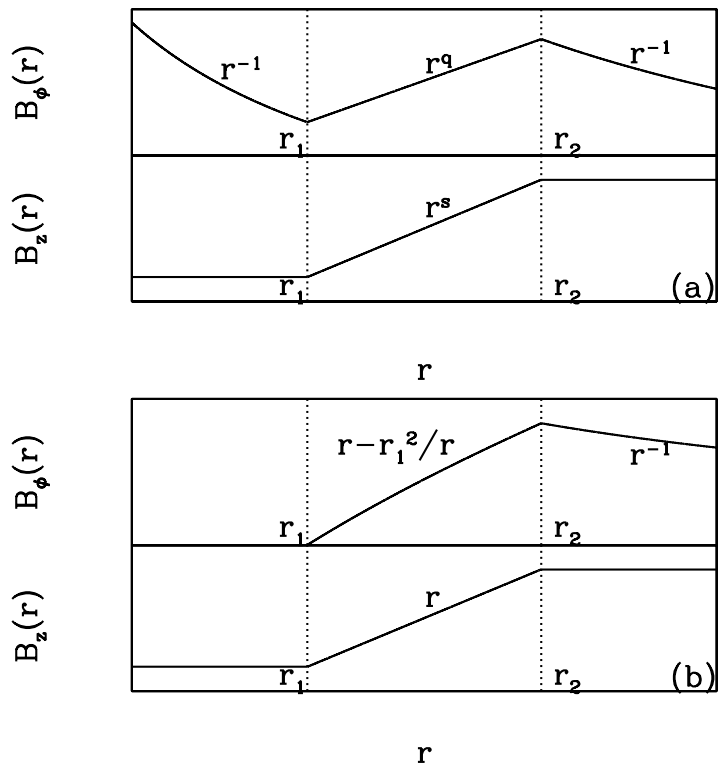}
\caption{The two magnetic field profiles adopted in the equilibrium torus model.}
\label{fig:1}
\end{center}
\end{figure}

In this model, we assume that there is an external current running vertically
at small radii much inside $r_1$, giving rise to
$B_{\phi}(r) \propto r^{-1}$ in the inner region ($r<r_1$).
There is no azimuthal current in this
region, so $B_z$ is constant. In the fluid zone, we adopt a
power-law magnetic field profile
\be
B_{\phi}(r) \propto r^{q},
\quad  B_z(r) \propto r^{s},
\ee
which means that both the azimuthal and vertical
current densities are also of power-law form.
Outside the fluid zone ($r > r_2$), there is no current.
Hence, $B_{\phi}(r) \propto r^{-1}, ~B_z(r)={\rm const}$. The complete magnetic
field profile is illustrated in the upper two panels of Fig.~\ref{fig:1}.

Integrating the radial equilibrium equation
\be
{1\over\rho}\frac{d\Pi}{dr}=-{GM\over r^2}+r\Omega^2-\frac{B_{\phi}^2}
{4\pi\rho r},\label{eq:5}
\ee
gives the gas pressure profile
\begin{equation}
\frac{P}{\rho}={GM\over r}+\frac{r^2\Omega^2}{2p+2}-\frac{1}{2}v_{Az}^2-\frac{1}{2}
\left( 1+\frac{1}{q}\right)v_{A\phi}^2-C,
\label{eq:6}
\end{equation}
where
\be
v_{Az}=B_{z}/\sqrt{4\pi \rho},\quad
v_{A\phi}=B_{\phi}/\sqrt{4\pi \rho}
\ee
are the \Alfven velocities and
$C$ is the integration constant.
The location of gas pressure maximum is determined by
\be
\frac{d}{dr}(P/\rho)=-\frac{GM}{r^2}-s\frac{v_{Az}^2}{r}-(1+q)\frac{v_{A\phi}^2}{r}+r\Omega^2=0,
\label{eq:7}
\ee
which defines a reference radius $r_0$:
\be
\frac{GM}{r_0^2}=r_0 \Omega_0^2-s\frac{v_{Az}^2(r_0)}{r_0}-(1+q)\frac{v_{A\phi}^2(r_0)}{r_0}.
\label{eq:8}
\ee
Let $C \equiv \lambda GM/r_0$ with $\lambda$ being a constant and use Eq.~(\ref{eq:8}) to
substitute $GM$ in Eq.~(\ref{eq:6}), we can rewrite
Eq.~(\ref{eq:6}) in the dimensionless form
\be
\frac{P}{\rho}=\frac{1}{r}-\lambda+\frac{r^{2p+2}}{2p+2}+sv_{Az0}^2\left(-\frac{1}{r}
+\lambda-\frac{1}{2s}r^{2s}\right)
+(1+q)v_{A\phi0}^2\left(-\frac{1}{r}+\lambda-\frac{1}{2q}r^{2q}\right),
\label{eq:9}
\ee
where
\be
v_{A\phi0}=v_{A\phi}(r_0)/(r_0\Omega_0),\quad
v_{Az0}=v_{Az}(r_0)/(r_0\Omega_0).
\ee
Here and hereafter we will use units such that $r_0=\Omega_0=1$.
Once we specify $p$, $q$, $s$, $v_{A\phi0}$, $v_{Az0}$ and $\lambda$, we can determine
the locations of the torus boundary by solving $P=0$.
However, there are several constraints on these parameters:

(i) $dP/dr=0$ only guarantees the extremum of the $P(r)$ profile.
To ensure that we find a pressure maximum instead of minimum, we require
$d^2P/dr^2 <0$ at $r=r_0=1$, which implies
\be
2p+3-s(1+2s)v_{Az0}^2-(1+q)(1+2q)v_{A\phi0}^2 <0.
\label{eq:10}
\ee
This requirement reduces to $p< -3/2$ in the $B=0$ limit.

(ii) Both sides of Eq.~(\ref{eq:8}) need to be positive so that the gas
pressure maximum exists. Thus
\be
1-sv_{Az0}^2-(1+q)v_{A\phi0}^2 >0.
\label{eq:11}
\ee

(iii) The maximum gas pressure $P_{\rm max}$ must be positive. Thus, requiring the
RHS of Eq.~(\ref{eq:9}) to be positive at $r=1$ gives
\be
\lambda < \frac{1}{1-v_{Az0}^2-2v_{A\phi0}^2}
\left[\frac{2p+3}{2p+2}-\frac{3}{2}v_{Az0}^2-3v_{A\phi0}^2\right],
\label{eq:12}
\ee
provided that Eq. (\ref{eq:11}) is satisfied.

\begin{figure}
\begin{center}
\includegraphics[scale=0.5]{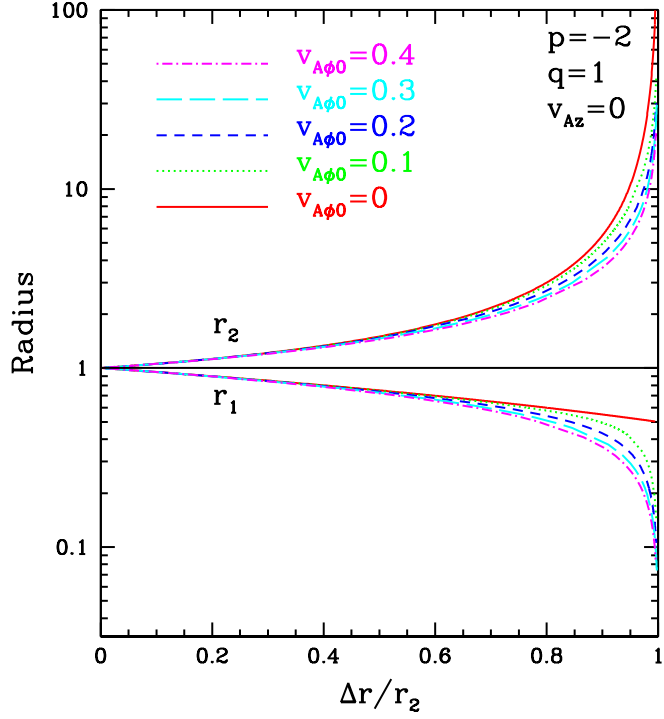}
\caption{Some examples of Model (a) with a pure toroidal magnetic
  field ($B_\phi\propto r$ in the fluid and $B_z=0$) and constant
  angular momentum distribution ($\Omega\propto r^{-2}$).  The x-axis
  is the relative thickness of the torus with $\Delta r=r_2-r_1$ being
  the absolute width, and the y-axis shows the locations of the two
  boundaries.  The different lines represent different values of
  $v_{A\phi 0}=v_{A\phi}(r_0)/(r_0\Omega_0)$, as indicated. The
  horizontal line indicates the location of gas pressure maximum
  ($r_0=1$).
}\label{fig:2}
\end{center}
\end{figure}

Figure~\ref{fig:2} illustrates some examples of Model (a).
We specify the values of $p$, $q$, $s$, $v_{A\phi0}$ and $v_{Az0}$,
then by varying $\lambda$, we obtain solutions for different torus thickness.
For a given $\Delta r/r_2$, both $r_1$ and
$r_2$ change when $v_{A\phi0}$ changes as a result of magnetic support
in the torus.
In the hydro limit ($v_{A\phi 0}=v_{Az0}=0$), $r_1$ approaches $0.5$
as $\Delta r/r_2 \rightarrow 1$. This feature
is shown analytically in Pringle \& King (2007).
For a finite field strength, we see that
$r_1 \rightarrow 0$ and $r_2 \rightarrow \infty$ as $\Delta r/r_2 \rightarrow 1$.
Note that for a relatively thin
torus ($\Delta r/r_2 \lesssim 0.6$), the differences of $r_2$ and $r_1$ between
different field strengths are quite small.
Since $B_{\phi}$ and $B_z$ have similar effects on the equilibrium
structure (see Eq.~[\ref{eq:9}]), these features also apply to models with a
finite vertical field.
The special case of $s=0$ and $q=-1$ is worth mentioning: In this case, the magnetic
field is force-free and has no effect on the equilibrium structure
\footnote{This is why in Curry \& Pudritz (1996) the one-to-one mapping
between $r_2/r_1$ and $(r_2-r_0)/r_2$ remains unchanged for different
uniform vertical B field strengths.}.

\subsection{Model (b)}

The magnetic field profile in this case is shown in the bottom two
panels of Fig.~\ref{fig:1}.  Compared with Model (a), the difference
is that there is no vertical current at small radius. Therefore
the azimuthal field in the inner region ($r <r_1$) is zero. In the
fluid zone, we assume $B_z \propto r$ and $B_{\phi}(r) \propto
r-r_1^2/r$ such that both azimuthal and vertical current densities are
uniform.  Following the same procedure as in section 2.1, we
can derive the dimensionless expression for gas pressure profile:
\be
\frac{P}{\rho}=\frac{1}{r}-\lambda+\frac{r^{2p+2}}{2p+2}+v_{Az0}^2
\left(\lambda-\frac{1}{r}-\frac{1}{2}r^2\right)
+v_{A\phi0}^2\left[\left(\lambda-\frac{1}{r}\right)\frac{1+r_1^2}{1-r_1^2}
-\frac{1}{(1-r_1^2)^2}(r^2-r_1^2-2r_1^2\ln r)\right].
\label{eq:13}
\ee
Similarly, for a viable equilibrium model to exist, the model parameters must satisfy
the following requirements:
\be
2p+3-3v_{Az0}^2-\frac{3+r_1^4}{(1-r_1^2)^2}v_{A\phi0}^2<0,
\label{eq:14}
\ee
\be
1-v_{Az0}^2-\frac{1+r_1^2}{1-r_1^2}v_{A\phi0}^2>0,
\label{eq:15}
\ee
\be
\lambda < \frac{1}{1-v_{Az0}^2-\frac{1+r_1^2}{1-r_1^2}v_{A\phi0}^2}\left(\frac{2p+3}{2p+2}
-\frac{3}{2}v_{Az0}^2-\frac{2+r_1^2}{1-r_1^2}v_{A\phi0}^2\right).
\label{eq:16}
\ee

%%%%%%%%%%%%%%%%%%%%%%%%%%%%%%%%%%%%%%%%%%%%%%%%%%%%%%
\section{MHD Equations for Perturbations}

Assuming that the Eulerian perturbation of any physical variable $f$ is of the
form $\delta f \propto e^{im\phi-i\omega t}$
(with no dependance on $z$), the linearized perturbation equations are
\ba
\frac{1}{r}\frac{\partial}{\partial r}(r\delta u_r)+\frac{im}{r}\delta u_{\phi}&=&0\label{eq:17}\\
-i\tomega\delta u_r-2\Omega\delta u_{\phi}&=&-\frac{1}{\rho}\frac{\partial \delta \Pi}{\partial r}
+\frac{imB_{\phi}}{4\pi\rho r} \delta B_r-\frac{B_{\phi}}{2\pi\rho r}\delta B_{\phi}\label{eq:18}\\
-i\tomega \delta u_{\phi}+\frac{\kappa^2}{2\Omega}\delta u_r&=&
-\frac{im}{\rho r}\delta \Pi +\frac{1}{4\pi\rho}\left(\frac{\partial}{\partial r}+\frac{1}{r}\right)B_{\phi}
\delta B_r+\frac{im B_{\phi}}{4\pi\rho r} \delta B_{\phi}\label{eq:19}\\
-i\tomega \delta u_z &=&\frac{im B_{\phi}}{4\pi\rho r}\delta B_z+\frac{1}{4\pi\rho}\frac{dB_z}{dr}
\delta B_r\label{eq:20}\\
-i\tomega \delta B_r&=&im \frac{B_{\phi}}{r}\delta u_r\label{eq:21}\\
-i\tomega \delta B_{\phi}&=&\frac{imB_{\phi}}{r}\delta u_{\phi}-r\frac{d}{dr}\left(\frac{B_{\phi}}{r}\right)\delta u_r
+r\frac{d\Omega}{dr}\delta B_r\label{eq:22}\\
-i\tomega \delta B_z&=&\frac{imB_{\phi}}{r}\delta u_z-\frac{d B_z}{dr}\delta u_r,\label{eq:23}
\ea
where $\tomega=\omega-m\Omega$ is the wave frequency in the
co-rotating frame and the radial epicyclic frequency $\kappa$ is given by
\be
\kappa^2=\frac{2\Omega}{r}\frac{d}{dr}(r^2\Omega)=2(p+2)\Omega^2.
\label{eq:24}
\ee
Using $\Delta \bb{u}=\delta
\bb{u}+\bb{\xi}\cdot\nabla\bb{u}=d\bb{\xi}/dt
=-i\omega\xi+(\bb{u}\cdot \nabla)\bb{\xi}$, we find that the
Eulerian perturbation $\delta \bb{u}$ is related to the Lagrangian
displacement vector $\bb{\xi}$ by $\delta
\bb{u}=-i\tomega\bb{\xi}-r\Omega'\xi_r \hat{\bb{\phi}}$ (prime donates
radial derivative) and
we can further combine Eqs.~(\ref{eq:17})-(\ref{eq:23}) into
two equations for $\xi_r$ (radial Lagrangian displacement) and $\delta \Pi /\rho$:
\ba
&&\frac{d\xi_r}{dr}=A_{11}\xi_r+A_{12}\frac{\delta \Pi}{\rho}, \label{eq:25}\\
&&\frac{d}{dr}\left(\frac{\delta \Pi}{\rho}\right)=A_{21}\xi_r+A_{22}\frac{\delta \Pi}{\rho},
\label{eq:26}
\ea
where
\ba
&&A_{11}=-\frac{1}{r}\frac{\tomega^2-2m\tomega\Omega+m^2\omega_{A\phi}^2}
{\tomega^2-m^2\omega_{A\phi}^2},
\label{eq:27} \\
&&A_{12}=\frac{m^2}{r^2},
\label{eq:28} \\
&&A_{21}=\tomega^2-m^2\omega_{A\phi}^2-2r\Omega\frac{d\Omega}{dr}
+\left(2\frac{d\ln B_{\phi}}{d\ln r}-1\right)
\omega_{A\phi}^2-4\frac{(\tomega\Omega+m\omega_{A\phi}^2)^2}{(\tomega^2-m^2\omega_{A\phi}^2)},
\label{eq:29} \\
&&A_{22}=\frac{2m}{r}\frac{\tomega\Omega+m\omega_{A\phi}^2}{\tomega^2-m^2\omega_{A\phi}^2},
\label{eq:30} \ea
and $\omega_{A\phi}\equiv v_{A\phi}/r=B_{\phi}/({r\sqrt{4\pi\rho}})$ is
the toroidal \Alfven frequency.
Equations (\ref{eq:25}) and (\ref{eq:26}) are the same as Eqs.~(119) and
(120) (derived for a pure toroidal magnetic field) in Chandrasekhar (1961).

Note that although we start with a mixed magnetic field
$\bb{B}=B_{\phi}\hat{\bb{\phi}}+B_z\hat{\bb{z}}$, the final
Eqs.~(\ref{eq:25}) and (\ref{eq:26}) do not contain $B_z$. The reason is that
the $z$-component only appears in Eqs.~(\ref{eq:20}) and
(\ref{eq:23}), which in fact can be decoupled from the other five
perturbation equations.
Indeed, using Eq.~(\ref{eq:21}) to replace
$\delta B_r$ in Eq.~(\ref{eq:20}), and combining with
Eq.~(\ref{eq:23}), we find
\be
(\tomega^2-m^2\omega_{A\phi}^2)
\left(\delta B_z+{i\over\tomega}{dB_z\over dr}\delta u_r\right)=0.
\ee
In general, $\tomega^2-m^2\omega_{A\phi}^2\neq 0$. Comparing the
above equation with Eq.~(\ref{eq:23}) we have $\delta u_z=0$.
This is to be expected since the perturbed quantities are assumed
to be independent of $z$.
Also note that when the wave frequency $\omega$ is real, the coefficients
$A_{11},\,A_{21}$ and $A_{22}$ are singular at
\be
\tomega^2=m^2\omega_{A\phi}^2. \label{eq:31}
\ee
We shall call them the Magnetic Resonances (MRs).
Obviously, they reduce to the corotation resonance
when $B_\phi=0$.

%%%%%%%%%%%%%%%%%%%%%%%%%%%%%%%%%%%%%%%%%%%%%%%%%
\section{Boundary Conditions}

In general, the boundary of any magnetized flow should satisfy the following
conditions:
\ba
\left[\rho u_n\right]&=&0, \label{eq:46}\\
\left[\bb{n}\cdot \bb{B}\right]&=&0, \label{eq:32}\\
\left[P+\rho u_n^2+\frac{B_t^2}{8\pi}\right]&=&0, \label{eq:33}\\
\left[\rho u_n \bb{u_t}-\frac{B_n \bb{B_t}}{4\pi}\right]&=&0,\label{eq:34}
\ea
where $\bb{n}$ is a unit vector normal to the boundary surface, the subscript $n$
and $t$ denote the normal and tangential components,
and the square bracket represents the difference in a quantity across the boundary
(e.g., Schmidt 1979; Shu 1992).
For the system we study in this paper, there is no radial background
flow, we only need to consider
Eqs.~(\ref{eq:32})-(\ref{eq:34}) with $u_n=0$.
Obviously, with the magnetic field continuous across the boundaries and
with no radial field component, our equilibrium models constructed in Sec.~2
already satisfy the boundary conditions.
In the perturbed state, the boundary conditions read
\ba
\Delta\left[\bb{n}\cdot\bb{B}\right]&=&0, \label{eq:35} \\
\Delta\left[P+\frac{B_t^2}{8\pi}\right]&=&0, \label{eq:36}\\
\Delta\left[\frac{B_n \bb{B_t}}{4\pi}\right]&=&0. \label{eq:37}
\ea
Note that $\Delta[\bb{n}\cdot \bb{B}]=
[(\Delta \bb{n})\cdot \bb{B}]+[\bb{n}\cdot\Delta \bb{B}]$,
and since both $\bb{n}$ and $\Delta \bb{n}$ are the same across the boundary
\footnote{Note that in the non-axisymmetric case,
the perturbed surface normal vector $\Delta\bb{n}$ is not the same as
${\hat{\bb{r}}}$;
see Schmidt (1979) for a derivation of $\Delta\bb{n}$.},
Eq.~(\ref{eq:35}) simply becomes
$\bb{n}\cdot\left[\Delta\bb{B}\right]=0$,
where we have used $[\bb{B}]=0$ (as assumed in our model setup).
Since
$(\Delta\bb{B})_{r}=\left(\delta \bb{B}+\bb{\xi}\cdot \nabla\bb{B}\right)_r
=\delta B_r-\xi_{\phi}B_{\phi}/r$,
and both $\xi_{\phi}$ and $B_{\phi}$ are continuous across the boundaries,
we find
\be
[\delta B_r]=0.
\label{eq:42}
\ee
On the other hand, Eqs.~(\ref{eq:35}) and (\ref{eq:36}) combine to give
\be
\left[\Delta \Pi\right]=0. \label{eq:40}
\ee
The condition (\ref{eq:37}) is already satisfied because
$\Delta \left[{B_n \bb{B_t}}\right]=
\left[(\Delta B_n)\bb{B_t}\right]
+\left[{B_n\Delta \bb{B_t}}\right]=0$ when the background field is continuous.
We note that
the perturbed magnetic field does not need to be continuous across the boundary.
This means that there could be surface current induced by the perturbation.

To implement the two boundary conditions (\ref{eq:42})-(\ref{eq:40}),
we need to calculate the perturbed magnetic field in the vacuum region
($r <r_1$ and $r> r_2$). This can be done by solving
\ba
\nabla \times \delta \bb{B} =0,\quad
\nabla \cdot \delta \bb{B}=0. \label{eq:53}
\ea
Clearly, $\delta \bb{B}$ is a potential field $\delta \bb{B}=\nabla \Psi$ with
$\Psi$ (also $\propto e^{im\phi-i\omega t}$) satisfying
\be
\nabla^2 \Psi=0.\label{eq:54}
\ee
The solution of Eq.~(\ref{eq:54}) is
\be
\Psi=C_1 r^m+C_2r^{-m}, \label{eq:55}
\ee
where $C_1$ and $C_2$ are integration constants.
Requiring $\delta \bb{B}$ to be regular at $r\rightarrow0$ and $r\rightarrow\infty$,
we find
\be
\delta B_r=C_1mr^{m-1}, ~\delta B_{\phi}=C_1\frac{im}{r}r^m,\quad\mbox{for}~r < r_1,
\label{eq:57}
\ee
and
\be
\delta B_r=-C_2mr^{-m-1}, ~\delta B_{\phi}=C_2\frac{im}{r}r^{-m}, \quad
\mbox{for}~r>r_2. \label{eq:58}
\ee
The two constants $C_1$ and $C_2$ can be determined by using $[\delta B_r]=0$.
At $r=r_1$, this implies
$C_1mr^{m-1}=imB_{\phi}\xi_r/r$
(see Eq.~[\ref{eq:21}] for $\delta B_r$ inside the fluid zone).
Thus,
\be
C_1=iB_{\phi}\xi_r r^{-m}|_{r_1}.\label{eq:60}
\ee
Similarly,
\be
C_2=-iB_{\phi}\xi_r r^m|_{r_2}. \label{eq:61}
\ee
Since the detailed realization of the boundary condition
$[\Delta\Pi]=0$ depends on the specific equilibrium
model, we address the two models separately.

\begin{figure}
\begin{center}
\includegraphics[scale=0.6]{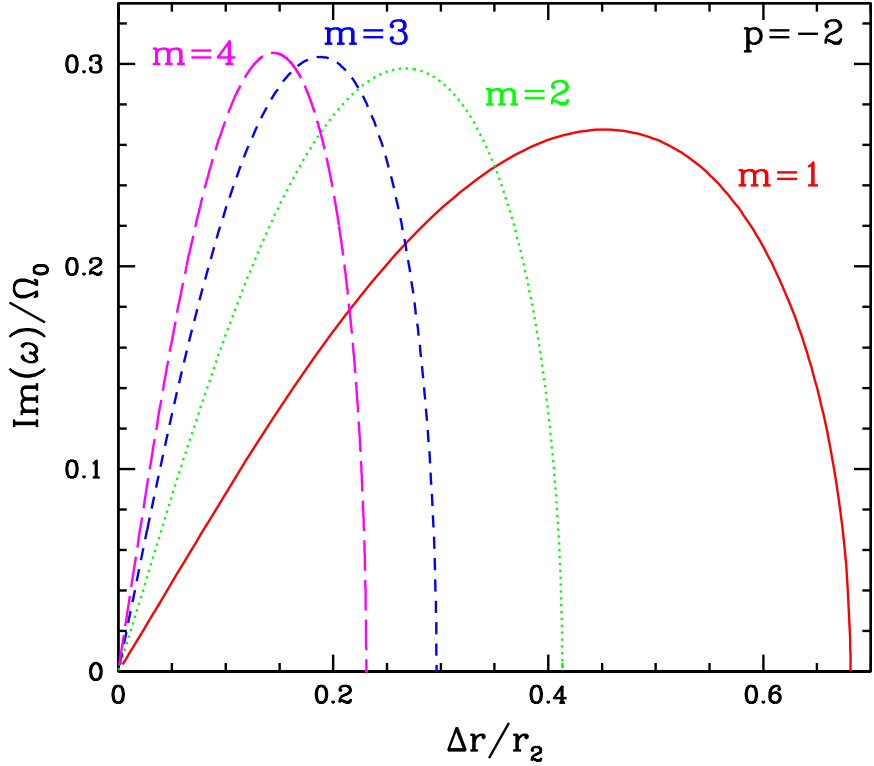}
\caption{The growth rate of Papaloizou-Pringle instability (in units of
$\Omega_0$, the fluid rotation rate at the pressure maximum of the torus)
as a function of the relative thickness of the torus for different values of $m$.
The rotation profile is $\Omega\propto r^{-2}$.
This figure is similar to Fig.~1 in Blaes \& Glatzel (1986)
and to Fig.~1 in Abramowicz et al.~(1987).}
\label{fig:3}
\end{center}
\end{figure}

\begin{figure}
\begin{center}
\includegraphics[scale=0.5]{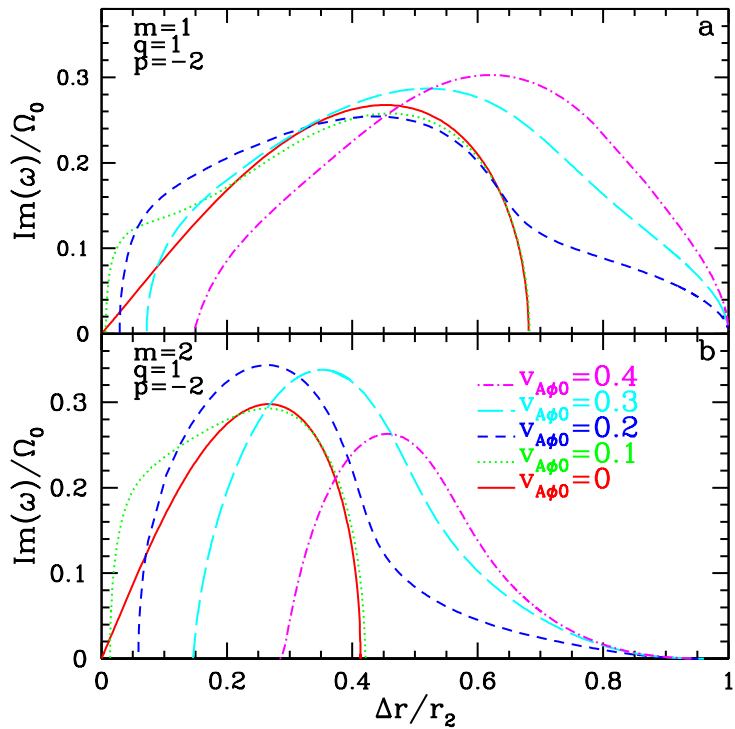}
\caption{The instability growth rate as a function of the relative thickness of
the torus for a range of toroidal magnetic field strengths. Different lines represent
different $v_{A\phi0}$
with the solid line denoting the hydrodynamic case. The upper and
bottom panels are for the $m=1$ and $m=2$ modes, respectively. }
\label{fig:4}
\end{center}
\end{figure}

\begin{figure}
\begin{center}
\includegraphics[scale=0.5]{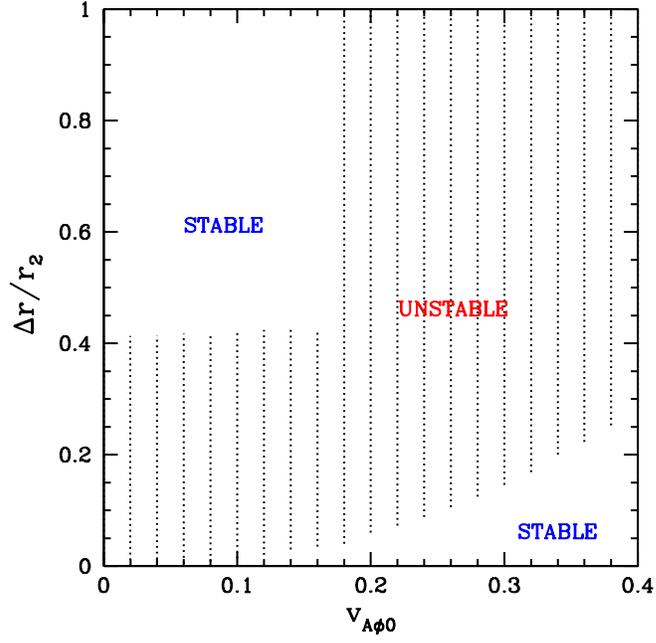}
\caption{The instability region in the parameter space defined by the dimensionless
torus thickness $\Delta r/r_2$ and
the toroidal magnetic field strength. The other parameters are fixed to $m=2$,
$\Omega\propto r^{-2}$ and $B_\phi\propto r$.
The dotted area denotes the region where a growing mode can be found.}
\label{fig:5}
\end{center}
\end{figure}

\begin{figure}
\begin{center}
\includegraphics[scale=0.5]{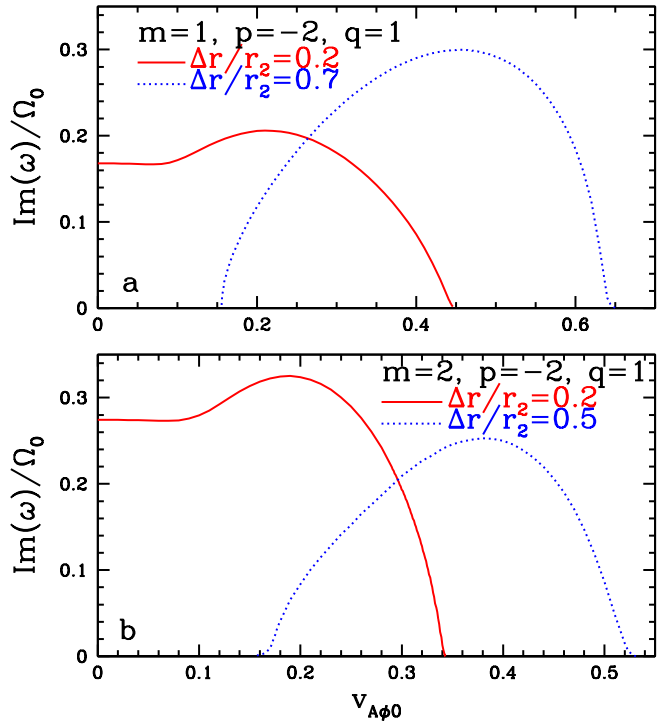}
\caption{The instability growth rate as a function of $v_{A\phi
    0}=v_{A\phi}(r_0)/(r_0\Omega_0)$ for tori with different thickness
  $\Delta r/r_2$. The upper and bottom panels depict the cases with
  $m=1$ and $m=2$, respectively.}
\label{fig:6}
\end{center}
\end{figure}

\begin{figure}
\begin{center}
\includegraphics[scale=0.5]{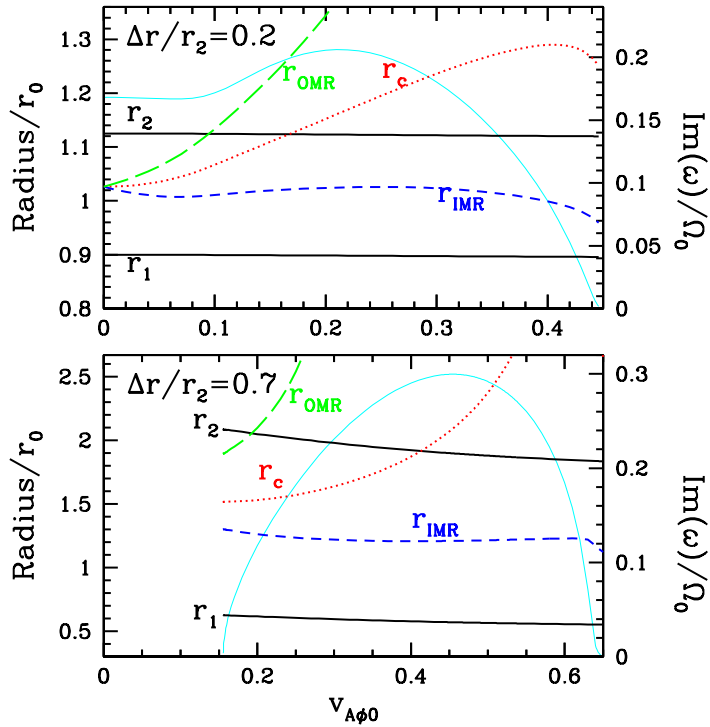}
\caption{Some special radii for the $m=1$ overstable mode in a torus
  as a function of the dimensionless magnetic field strength
  $v_{A\phi0}$. The upper and bottom panels correspond to a thin
  $\Delta r/r_2=0.2$) and thick ($\Delta r/r_2=0.7$) torus,
  respectively.  The two solid lines show the inner and outer torus
  boundaries ($r_1$ and $r_2$). The dotted line represents the
  corotation radius, while the shot-dashed and long-dashed lines show
  the inner and outer magnetic resonances (IMR and OMR), respectively
  [see Eq.~(\ref{eq:mr})].  The thin solid line is the mode growth
  rate with the scale shown on the right.  The torus rotation and
  magnetic field profiles are $\Omega\propto r^{-2}$ and
  $B_\phi\propto r$.}
\label{fig:7}
\end{center}
\end{figure}

\begin{figure}
\begin{center}
\includegraphics[scale=0.5]{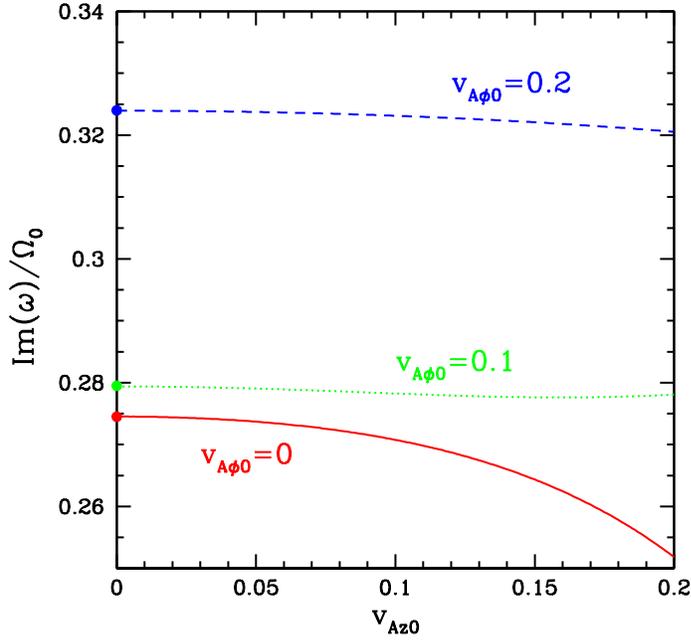}
\caption{The instability growth rate as a function of $v_{Az0}$ for a torus
with relative thickness $\Delta r/r_2=0.2$.  Different lines
represent different values of $v_{A\phi0}$. The other parameters are
fixed to $m=2$, $\Omega\propto r^{-2}$, $B_\phi\propto r$ and
$B_z\propto r$ in the fluid.}
\label{fig:9}
\end{center}
\end{figure}

\begin{figure}
\begin{center}
\includegraphics[scale=0.5]{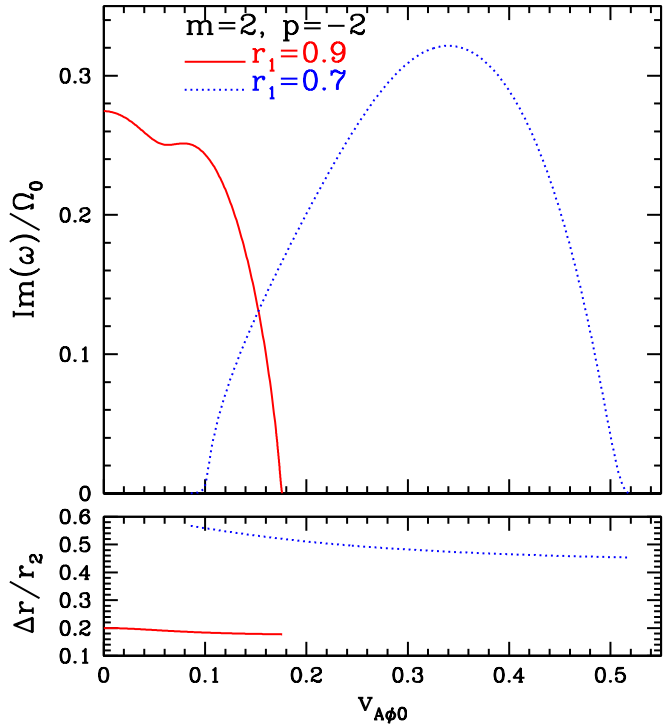}
\caption{The instability growth rate as a function of $v_{A\phi0}$ for
accretion tori described by Model (b) with a pure toroidal magnetic field.
The solid and dotted lines correspond to two different inner disc boundary
radii. The bottom pane shows the corresponding torus relative thickness.}
\label{fig:10}
\end{center}
\end{figure}

\begin{figure}
\begin{center}
\includegraphics[scale=0.5]{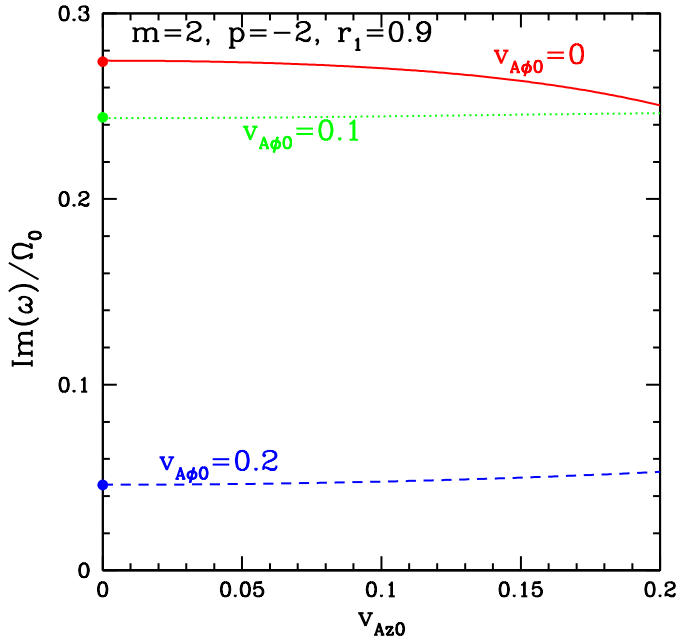}
\caption{The instability growth rate as a function of $v_{Az0}$ for a thin torus
(with $r_1=0.9$ and $\Delta r/r_2\simeq 0.2$).
The different lines represent different values of $v_{A\phi0}$.
Unlike Fig.~\ref{fig:9}, here the magnetic field profile is described by
Model (b).}
\label{fig:11}
\end{center}
\end{figure}

%%%%%%%%%%%%%%%%%%%%%%%%%%%%%%%%%%%%%
\subsection{Model (a)}

At the inner boundary $r=r_1$, by using
Eq.~(\ref{eq:60}),
we find that the Lagrangian perturbation of the total
pressure in the vacuum just inside $r_1$ (i.e., $r=r_{1^-}$) is given by
\be
\Delta \Pi|_{r_{1^-}}=\frac{B_{\phi}
\delta B_{\phi}}
{4\pi}-\xi_r\frac{B_{\phi}^2}{4\pi r}
=-(m+1)\xi_r\frac{B_{\phi}^2}{4\pi r}. \label{eq:64}
\ee
In the fluid just outside $r_1$ (i.e., $r=r_{1+}$), we have
\be
\Delta\Pi|_{r_{1^+}}=\delta\Pi+\xi_r\left[r^{2p+1}-\frac{v_{A\phi}^2}{r}
-\frac{1}{r^2}\left(1-sv_{Az0}^2-(1+q)v_{A\phi0}^2\right)\right]\rho.
\label{eq:65}
\ee
Thus the condition $[\Delta \Pi]=0$ at $r=r_1$ can be written as
\be
\frac{\delta \Pi}{\rho}+\xi_r\left[r^{2p+1}+
{mv_{A\phi}^2\over r}
-\frac{1}{r^2}\left(1-sv_{Az0}^2-(1+q)v_{A\phi0}^2\right)\right]=0,
\quad{\rm at}~r=r_{1^+}.
\label{eq:66}
\ee
Note that in deriving the above equation, we have implicitly used $[B_{\phi}]=0$
and $[\xi_r]=0$. The same
procedure yields the boundary condition at $r=r_2$
\be
\frac{\delta \Pi}{\rho}+\xi_r\left[r^{2p+1}
-{mv_{A\phi}^2\over r}
-\frac{1}{r^2}
\left(1-sv_{Az0}^2-(1+q)v_{A\phi0}^2\right)\right]=0, \quad{\rm at}~r=r_{2^-}.
\label{eq:67}
\ee

\subsection{Model (b)}

The derivation is similar to Model (a). In this case, the boundary conditions are
\be
\frac{\delta \Pi}{\rho}+\xi_r\left[r^{2p+1}-\frac{1}{r^2}
\left(1-v_{Az0}^2-\frac{1+r_1^2}{1-r_1^2}v_{A\phi0}^2\right)\right]=0,
\quad {\rm at}~r=r_{1^+},
\label{eq:68}
\ee
\be
\frac{\delta \Pi}{\rho}+\xi_r\left[r^{2p+1}-\frac{mv_{A\phi0}^2}{(1-r_1^2)^2}
\left(1-\frac{r_1^2}{r^2}\right)
-\frac{1}{r^2}
\left(1-v_{Az0}^2-\frac{1+r_1^2}{1-r_1^2}v_{A\phi0}^2\right)\right]=0,
\quad {\rm at}~r=r_{2^-}.
\label{eq:69}
\ee

%%%%%%%%%%%%%%%%%%%%%%%%%%%%%%%%%%%%%%%%%%%%%%%%%
\section{Numerical Results}

We employ the standard shooting method (Press et al.~1992) to solve
Eqs.~(\ref{eq:25}) and (\ref{eq:26}) subjected to the boundary
conditions (\ref{eq:66})-(\ref{eq:67}) [Model (a)] or
(\ref{eq:68})-(\ref{eq:69}) [Model (b)] to obtain the eigenvalue
$\omega=\omega_r+i\omega_i$. For most of our analysis, we set the
rotation index $p=-2$ (i.e., $\Omega\propto r^{-2}$)
such that our results can be directly compared
with the original Papaloizou-Pringle instability.

Before discussing our results for finite magnetic fields,
we briefly review the main features of the classical (hydrodynamical)
PP instability. As seen in Fig.~\ref{fig:3},
the instability growth rate increases with increasing torus thickness
for small $\Delta r/r_2$ but
terminates at some finite thickness. As $m$ increases, the
termination point shifts to smaller $\Delta r/r_2$, although the peak
growth rate remains approximately the same.
This means that the PP instability only exists for relatively thin tori, as
shown by Blaes \& Glatzel (1986) and by
Abramowicz et al.~(1987).
The former also provides an approximate analytical expression for the
limiting maximum growth rate as $m \rightarrow \infty$.
These features can be understood from the fact that
the PP instability arises from the interaction of the
surface gravity waves at the torus boundary.
For a thin torus, the velocity shear across the corotation point is
small, and there would not be enough shear rotational energy available to drive the
growth. When the torus thickness is too large,
the wave amplitudes at the corotation radius (where the waves
exchange angular momentum) is too small to allow for adequate interactions.
Thus, only for the ``intermediate'' torus thickness, with
$m\Delta r/r \lo 1$, will the instability operates.
This explains why
for a larger $m$, the PP instability terminates at a smaller torus thickness.

\subsection{Model (a): Pure Toroidal Field Configuration}

In this section, we present the numerical results for an accretion
torus with a pure power-law profile toroidal magnetic field. We choose
the power-law index $q=1$ so that the vertical current density is
uniform as in Model (b).

Figure \ref{fig:4} shows the growth rate $\omega_i$ as a function of
thickness $\Delta r/r_2$ for different $v_{A\phi0}$. The two panels
share similar characteristics: (i) For relatively weak B field
($v_{A\phi0} \lesssim 0.1$), the instability resembles the $B=0$ limit
in that it always starts from infinitely small thickness and
terminates beyond a certain $\Delta r/r_2$; (ii) For stronger B
fields, the instability starts beyond certain finite $\Delta r/r_2$
and then extends all the way to very large thickness (although as
$\Delta r/r_2$ approaches unity, the growth rate becomes increasingly
small).  As $v_{A\phi0}$ increases, the critical thickness for the
onset of instability also increases.

Figure~\ref{fig:5} maps out the unstable zone in the thickness --
magnetic field strength parameter space. It shows the similar feature
as as Fig.~\ref{fig:4}. We can see that the unstable region is mainly
located at the lower-left (thin torus with weak B field) and the
upper-right (thicker torus with strong B field) corners of the
parameter space.

In Fig.~\ref{fig:6}, we present our numerical results in a different
way. We fix the dimensionless thickness $\Delta r/r_2$ and plot the
growth rate as a function of magnetic field strength. For a thin torus, we
find that as $v_{A\phi0}$ increases, the growth rate first goes up
slightly compared to the $B=0$ case, then decreases and becomes
completely suppressed when the magnetic field is sufficiently strong
($v_{A\phi}$ comparable to rotation velocity).
For a thick torus, the instability can survive
for a relatively stronger B field and then vanishes beyond a certain
$v_{A\phi 0}$.

To probe the underlying physics of how magnetic fields
affect the PP instability, we show in Fig.~\ref{fig:7}
the locations of several special points in the fluid:
The corotation radius $r_{\rm c}$ is where the wave pattern corotates with
with the background flow, i.e., $\tomega_r=\omega_r-m\Omega=0$.
The inner/outer magnetic resonances (IMR/OMR) are
defined by [see Eq.~(\ref{eq:31})]
\be
\tomega_r=\omega_r-m\Omega=\pm m\omega_{A\phi}.
\label{eq:mr}\ee
At the IMR, the wave is trailing the background flow but
corotates with the azimuthal \Alfven wave traveling in the
counter-rotational direction (viewed in the corotating frame),
while at the OMR, the wave is leading the background flow and corotates
with the \Alfven wave in the rotational direction.
Recall that for PP instability to operate in the $B=0$ limit, it is
essential that the corotation radius lies in between torus boundaries
(i.e., $r_1<r_c<r_2$). Now, with the inclusion of the magnetic field,
we see from Fig.~\ref{fig:7} that as $v_{A\phi0}$ increases, both
$r_c$ and $r_{\rm OMR}$ shift beyond the outer boundary of the torus.
The IMR radius, $r_{\rm IMR}$, however, always stays inside the fluid.
This suggests that in a magnetic torus, the IMR plays a similar role
as the corotation resonance does in a non-magnetic torus.

%%%%%%%%%%%%%%%%%%%%%%%%%%%%%%%%%
\subsection{Other Magnetic Field Configurations}

\subsubsection{Model (a): Mixed magnetic field}

Although the vertical magnetic field $B_z$ does not enter into the
perturbation equations, the presence of a finite $B_z$ can affect the
mode growth rate through the boundary conditions.
Figure~\ref{fig:9} shows some results for the accretion tori with a
mixture of vertical and toroidal magnetic fields. We take the
power-law index of $B_z$ to be $s=1$ so that the azimuthal electric
current density is constant. In Fig.~\ref{fig:9}, we fix the toroidal
field and plot the mode growth rate as a function of the vertical
field strength. We find that the effect of finite $v_{Az0}$ is small
(note the scale of the y-axis). For $v_{Az}=0$, the results
agree with what is shown in Fig.~\ref{fig:6}b.

\subsubsection{Model (b): Pure toroidal field}

In this case, since the background toroidal magnetic field has a
profile that depends on the inner boundary radius $r_1$, to solve for
the equilibrium structure and the global mode, we must first specify
$r_1$. Once we fix $r_1$, we can easily solve for the other boundary
radius $r_2$.  In Fig.~\ref{fig:10}, we show in the upper panel the
mode growth rate as a function of $v_{A\phi 0}$. The result is
qualitatively similar to Fig.~\ref{fig:6}b.  The bottom panel shows
that for a fixed inner boundary radius $r_1$, the thickness does not change
appreciably as $v_{A\phi0}$ varies. So the two values of $r_1$ we
choose adequately depict the thin and thick tori, respectively. Again,
we see that for a thin torus, the original Papaloizou-Pringle
instability is suppressed by the toroidal field, while for a
thick torus, the instability can survive for larger field strengths.

\subsubsection{Model (b): Mixed magnetic field}

In Fig.~\ref{fig:11}, we show the mode growth rate as a function of
$v_{Az0}$ for different fixed values of $v_{A\phi0}$.  Similar to the
case shown in Fig.~\ref{fig:9}, we see that the vertical magnetic
field has a small effect on the stability property of a magnetized
accretion torus. Again, this is understandable given that $B_z$ does
not enter into the differential equations for the perturbations, but
only affects the modes through boundary conditions.

%%%%%%%%%%%%%%%%%%%%%%%%%%%%%%%%%%%%%%%%
\section{Discussion}

In this paper, we have studied the effect of magnetic fields
on the global non-axisymmetric instability (the PP instability)
in accretion tori. For simplicity, we assume that both the perturbation
and the background flow variables have no $z$-dependance (thus our tori
are essentially 2D cylinders). We have explored various possible
magnetic configurations in the torus.
Although the detailed property of the instability is model-dependent,
Figs.~\ref{fig:4}-\ref{fig:6} illustrate our general findings:
(i) For thin tori (with the dimensionless thickness
$\Delta r/r_2\lo 0.2$, where $r_2$ is the outer torus radius),
the instability exists for zero and weak magnetic fields, but is suppressed
when the toroidal field becomes sufficiently strong (with the corresponding
\Alfven speed $v_{A\phi}\go 0.2 r\Omega$ measured at gas pressure maximum);
(ii) For thicker tori ($\Delta r/r_2\go 0.4$),
the PP instability does not operate for zero and weak magnetic fields,
but becomes active when the field is sufficiently strong
($v_{A\phi}\go 0.2 r\Omega$ measured at gas pressure maximum).
A vertical magnetic field also influences the PP instability, but its effect
is generally smaller that that of the toroidal field.

It is difficult to precisely pin down the physical origin of the
magnetic field effect on the PP instability. For example, with a
finite toroidal magnetic field, we find that the corotation resonance
radius may lie outside the torus body, and yet the torus is still
unstable. On the other hand, the inner magnetic resonance radius,
where $\omega_r-m\Omega=- m\omega_{A\phi}$ [see Eq.~(\ref{eq:mr})],
always lie inside the fluid body. Thus we suspect that in a magnetic torus,
the inner magnetic resonance plays a similar role
as the corotation resonance does in a non-magnetic torus.

We note that the PP instability (or its magnetic generalization)
involves wave modes that do not have vertical structure
(i.e. $k_z=0$).  Thus it is distinctly different from the usual MRI
(Balbus \& Hawley 1998). Our finding about the instability of think
tori with strong magnetic fields is particularly interesting: Since
the MRI can be suppressed when the magnetic field is too strong, our
results suggest that magnetized tori may be subject to the instability
even when it is stable against the usual MRI.

\section*{Acknowledgments}
We thank Thierry Foglizzo, Omer Blaes and David Tsang for useful discussion during
the course of this study. This work has been supported in part by NASA
Grant NNX07AG81G and NSF grants AST 0707628. DL also acknowledges the
hospitality (January - June, 2010) of the Kavli Institute for
Theoretical Physics at UCSB, funded by the NSF through grant
PHY05-51164.

%%%%%%%%%%%%%%%%%%%%%%%%%%%%%%%%%%%%%%%%

\label{lastpage}
\end{document}